\documentclass{JCIN18}

\usepackage{framed}
\usepackage{lipsum}
\usepackage{mathtools}
\usepackage{cuted}
\usepackage{soul}

\begin{document}

\captionsetup[figure]{labelfont={bf},name={Fig.},labelsep=period}

\ArticleType{Research paper}
\Year{2018}
\Acknowl{The work was supported by Hong Kong Research Grants Council under the Grants 17208319, 17209917 and 17259416.}

\title{An Overview of Data-Importance Aware Radio Resource Management for Edge Machine Learning}

\author[1]{Dingzhu Wen}
\author[1]{Xiaoyang Li}
\author[1]{Qunsong Zeng}
\author[2]{Jinke Ren}
\author[1]{and Kaibin Huang}

\address[1]{Department of Electrical and Electronic Engineering, The University of Hong Kong, Hong Kong, China}
\address[2]{College of Information Science and Electronic Engineering, Zhejiang University, Hangzhou, China}


\abstract{The 5G network connecting billions of \emph{Internet-of-Things} (IoT) devices will make it possible to harvest an enormous amount of real-time mobile data. Furthermore, the 5G virtualization architecture will enable cloud computing at the (network) edge. The availability of both rich data and computation power at the edge has motivated Internet companies to deploy \emph{artificial intelligence} (AI) there, creating the hot area of edge-AI. Edge learning, the theme of this project, concerns training edge-AI models, which endow on IoT devices intelligence for responding to real-time events. However, the transmission of high-dimensional data from many edge devices to servers can result in excessive communication latency, creating a bottleneck for edge learning. Traditional wireless techniques deigned for only radio access are ineffective in tackling the challenge.
Attempts to overcome the communication bottleneck has led to the development of a new class of techniques for intelligent \emph{radio resource management} (RRM), called data-importance aware RRM. Their designs feature the interplay of active machine learning and wireless communication. Specifically, the metrics that measure data importance in active learning (e.g., classification uncertainty and data diversity) are applied to RRM for efficient acquisition of distributed data in wireless networks to train AI models at servers. This article aims at providing an introduction to the emerging area of importance-aware RRM. To this end, we will introduce the design principles, survey recent advancements in the area, discuss some design examples, and suggest some promising research opportunities.
}

\keywords{Radio resource management, scheduling, retransmission, edge machine learning, active learning}

\maketitle


\section{Introduction}
The traffic in mobile Internet is growing at a breathtaking rate due to the extreme popularity of mobile devices (e.g., smartphones and sensors). Analysis shows that there will be 80 billions of devices connected to Internet by 2025, resulting in a tenfold traffic growth compared with 2016 \cite{R1}. The availability of enormous mobile data and recent breakthroughs in \emph{artificial intelligence} (AI) motivate researchers to develop AI technologies at the network edge. Such technologies are collectively called edge AI and represent a latest trend in machine learning. {\bf Edge machine learning} concerns training of edge-AI models and is the theme of this paper \cite{R2,R3,R4A}. The migration of learning from central clouds towards the edge allows edge servers to have fast access to big data generated by edge devices in real time for fast training of AI models. In return, downloading the trained models to the devices provisions them intelligence to respond to real-time events. Presently, edge learning is the R\&D focus of leading Internet and telecommunication companies (e.g., Microsoft and Qualcomm) as they strive to apply AI to support \emph{Internet-of-Tings} (IoT) applications ranging from auto-pilot to healthcare.

Research on edge learning is cross-disciplinary as it merges two areas: computer science and wireless communication. They concern two key aspects of edge learning, namely \emph{fast data processing} and \emph{fast data acquisition} from edge devices (smartphones and sensors), respectively. The two aspects cannot be decoupled as their performances are interwound under a common goal of fast learning. While computing speeds are growing rapidly, wireless data acquisition suffers from scarcity of radio resources and hostility of wireless channels, resulting in a bottleneck for fast edge learning \cite{R4,R5}. In particular, edge learning typically requires uploading enormous data generated by a large number of edge devices. One example is the training of  Tesla's AI model for auto-driving using sensing data uploaded by millions of Tesla vehicles on the road. Given the conflict between large-scale data and scarce radio resources, wireless data acquisition can potentially cause excessive communication latency \cite{R6,R7,R8}. Therefore, suppressing communication latency poses a grand challenge for data acquisition in edge learning and calls for innovations on \emph{radio resource management} (RRM). 

Attempts to tackle this challenge has led to the emergence of a new class of RRM techniques to enable intelligent data acquisition for edge learning. They are based on the principle that radio resources are allocated to edge devices not only based on channel states (as in the conventional approach \cite{R9}) but also considering how important their data is for learning. This gives the name \emph{importance-aware RRM}. Conventional designs target radio access and thus focus on data rates \cite{R10} or \emph{Quality-of-Service} (QoS) \cite{R11}. In contrast, importance-aware RRM targeting edge learning will attempt to improve learning performance (i.e., learning speed and accuracy). The new objective calls for a new design approach integrating tools from machine learning and wireless communication. The aim of this article is to provide an overview of the area of importance-aware RRM by introducing the background, the new deign principles, and the research opportunities as well as recent advancements in different directions.



\section{Background}
\subsection{Wireless Communication and Machine Learning}
\begin{figure*}[h]
\begin{center}
{\includegraphics[width=12cm]{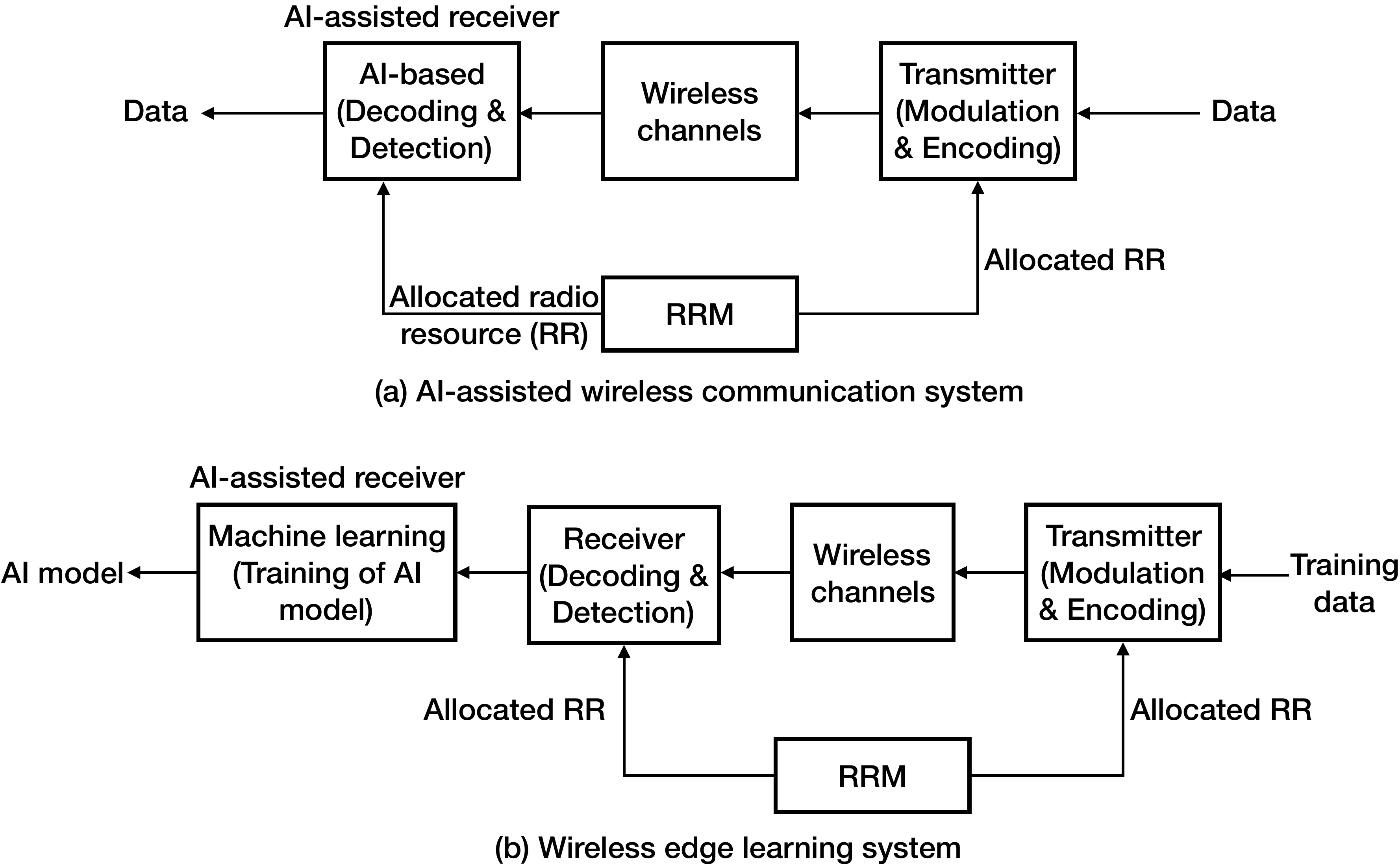}}\\
\caption{Comparison between (a) AI-comm system and (b) edge learning system.}
\vspace{-10pt} 
\label{fig:1}
\end{center}
\end{figure*}

Recent years have witnessed growing cross-disciplinary research merging wireless communication and machine learning. One research focus is to apply AI to improve the efficiency, robustness, and adaptivity of a communication system. Research in this vein forms the area of AI-assisted  wireless communication (AI-comm) where communication is the goal and learning is a tool. The emerging area of edge learning is fundamentally different where learning is the goal and wireless communication is the tool. Their differences in system operations are illustrated in Fig. \ref{fig:1}. AI-comm is introduced and compared with edge learning as follows.

\subsubsection{AI-assisted Wireless Communication (AI-comm)}
Traditional communication systems are based on a highly modular architecture divided into numerous functional blocks such as encoder/decoder and channel estimator. The resultant decomposed design not only compromises system performance but also results in high overhead and the lack of flexibility. Furthermore, existing system designs tend to assume oversimplified channel models. Consequently, theoretical performance is difficult to be fully materialized in practice. AI-comm research represents an attempt to overcome the above weaknesses by designing more integrated and versatile systems leveraging the power of AI \cite{R12,R13,R14,R15,R16}. In \cite{R12,R13}, researchers proposed novel schemes integrating channel estimation and data detection by applying sparse Bayesian learning. The innovation makes it possible to simultaneously recover a channel matrix and transmit symbols from the received signal with less channel-training overhead. Real channels are complex and frequently have no known models. To overcome the difficulty, researchers proposed in \cite{R14,R15} that deep learning can be applied to fit auto-encoders to a blackbox enclosing the channel and encoders/decoders. The deployment of such auto-encoders at transceivers provides a system adaptivity to an unknown propagation environment. Most recently, there have been breakthroughs in using recurrent-neural-network to construct a new class of error-correction codes \cite{R16}. They were shown to outperform even the state-of-the-art codes.

\subsubsection{AI-Comm vs. Edge Learning}
The main differences between AI-Comm and edge learning are elaborated as follows.
\begin{itemize}
\item {\bf Goal: communication vs. learning:} Typical AI-comm and edge-learning systems are compared in Fig. \ref{fig:1}. The goal of the former is communication, namely transportation of generic data from one place to another over the air. Rate maximization and reliability are the main design criteria. In contrast, the goal of an edge-learning system is learning, namely training of an AI model using distributed data. The main design criteria are learning speed and accuracy. Reliability may not be critical. For instance, for federated edge learning (see Section 2.2.2), perturbing stochastic gradients with noise can help avoid model overfitting or its being trapped at a local minimum \cite{R17,R18}. 

\item {\bf Passive vs. active learning:} For AI-Comm, the learning of an AI model (e.g., auto-encoder) is decoupled from communication and based on a given historical dataset from past  measurement. The learning process is \emph{passive} in the sense of lacking proactive data selection. In contrast, as elaborated in the next section, a key operation in edge learning is \emph{active data acquisition}, namely selecting and transmitting important data to accelerate the learning process, which is a main theme of this work and discussed in the next section. In other words, the edge-learning process is \emph{active} and integrated with communication.  

\end{itemize}

\subsection{Communication-Efficient Edge Learning}
\begin{figure*}[h]
\begin{center}
{\includegraphics[width=12cm]{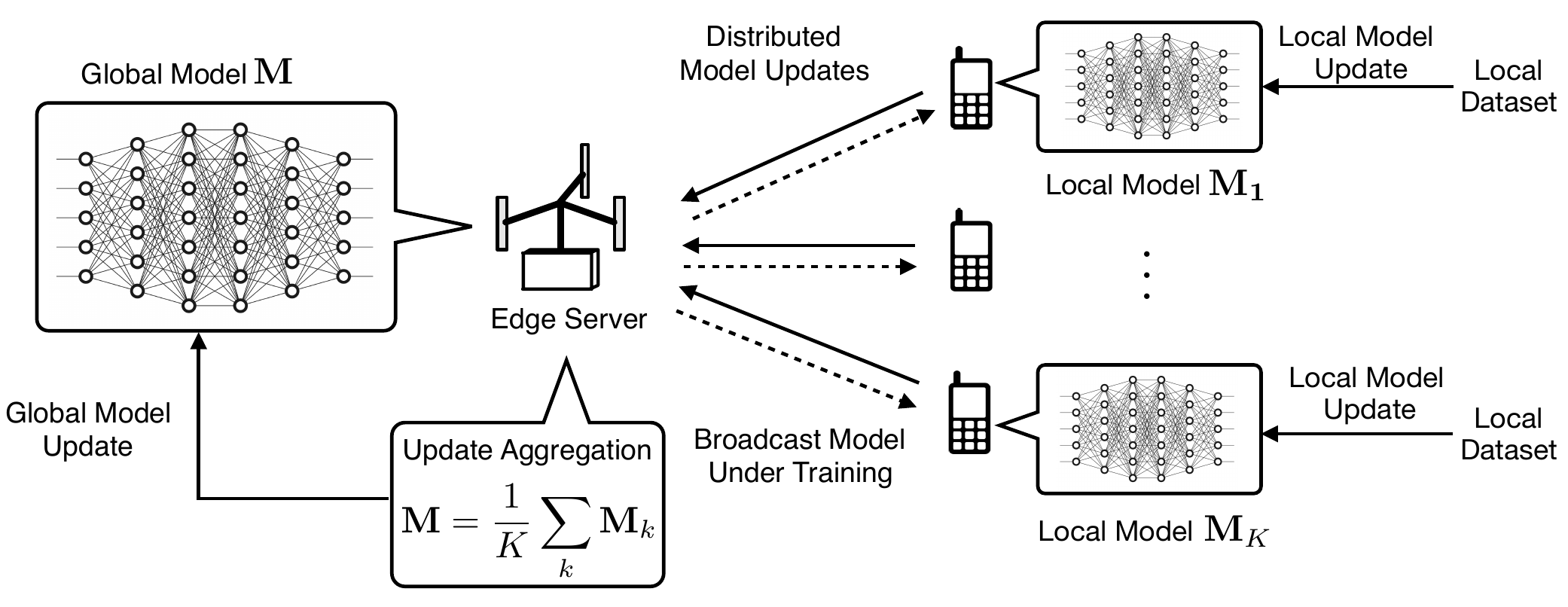}}\\
\caption{Federate edge learning system.}
\vspace{-10pt} 
\label{fig:2}
\end{center}
\end{figure*}

Distributed learning refers to machine learning at a server using data distributed at remote devices. Edge learning is a specific scenario of distributed learning with an air interface between server and devices. In traditional distributed learning studied mostly by computer scientists, communication channels are modelled coarsely as ``bit pipes" [4, 5]. Edge-learning research advances the area by the deployment of advanced communication techniques (e.g., RRM) for efficient data acquisition. There exist two paradigms for edge learning. In one paradigm called \emph{centralized edge learning} [see Fig. \ref{fig:1}(b)], learning is performed only at the server and training data is directly acquired from edge devices by wireless transmission \cite{R3}. To preserve user privacy, the other paradigm, called \emph{federated edge learning} (see Fig. \ref{fig:2}), avoids data uploading by distributing learning at both servers and edge devices, which is coordinated using wireless links \cite{R4,R19,R20}. Specifically, devices transmit to the server their updates on local AI models that are \emph{averaged} (see Fig. \ref{fig:2}) and then applied to update the global model. In view of high dimensionality in data/updates, the approach of importance-aware RRM has the key feature of \emph{active data/update acquisition} for communication-efficient edge learning. Its basic principle is to adopt notions of \emph{data importance} from the existing literature of machine learning for RRM. The notions are introduced separately for the two mentioned paradigms as follows.

\subsubsection{Centralized Edge Learning}
Consider distributed learning by direct data acquisition. To improve communication efficiency, it is natural to acquire only data samples that are ``important" for learning. Two data-importance metrics from the area of active learning are \emph{data uncertainty} and \emph{data diversity} \cite{R21}. Data classification is an important topic in machine learning that has a wide-range of applications ranging from computer vision to information retrieval. Targeting classification, active learning largely concerns the selection of data samples from a large unlabelled dataset for manual labelling and then training a classifier model. From the perspective of a single data sample, its informativeness for learning can be quantified by data (classification) uncertainty, referring to how confident it can be correctly classified using the real-time model \cite{R22}. On the other hand, when considering a batch of samples, the total informativeness depends not only on uncertainty of individual samples but also on their diversity, referring to non-overlapping information \cite{R23}.

Existing work on active learning by computer scientists focuses on data selection where communication is irrelevant. On the other hand, for importance-aware RRM research, communication engineers adopt the importance metrics from the area of active learning, namely data uncertainty and diversity, to design importance-aware RMM for \emph{active data acquisition} in edge learning.

\subsubsection{Federated Edge Learning}
There exist two schemes for federated learning featuring different types of update by edge devices. For one scheme, the edge devices transmit to the server their local models whose average replaces the global model. The other scheme is based the method of \emph{stochastic-gradient descent} (SGD), where devices transmit local gradient vectors and their average is applied to update the global model using SGD. For either scheme, the learning procedure iterates between 1) the broadcast of the global model by the server for updating local models and 2) transmission by edge devices for updating the global model. The iteration continues until the global model converges. Each iteration is called a \emph{communication round}. For federated learning, the concern of communication overhead has also motivated researchers to develop algorithms for selecting only a subset of devices with important updates for transmission. The theme is called \emph{active (model) update acquisition}. There exist two importance metrics corresponding to the two mentioned schemes. One is \emph{model variance} which indicates the divergence of a particular local model to the average of all local models. The other is \emph{gradient divergence} that reflects the level of changes on the current gradient update w.r.t. the previous one. Using these metrics to schedule updating devices, a number of so called ``lazily updating" algorithms have been designed for communication-efficient federated learning \cite{R24, R25}. Another approach for reducing communication overhead is to compress gradient vectors by exploiting their sparsity in significant elements \cite{R26, R27}. 

Importance-aware RRM in the context of federated edge learning builds on the mentioned update-importance metrics. Thereby, data importance analytics provides a new dimension for improving the communication efficiency and coping with the negative effect of wireless propagation on learning performance.

\subsubsection{Radio Resource Management}
Radio resources, primarily referring to time, frequency and space, create signalling dimensions, which are used in cellular systems to provide radio access to mobile users \cite{R10, R28}. Dividing and sharing of each type of resource among users have resulted in a set of multi-access schemes including \emph{time-division multi-access} (TDMA) \cite{R29}, \emph{orthogonal frequency-division multi-access} (OFDMA) \cite{R30, R31}, \emph{space-division multi-access} (SDMA) \cite{R32}, and \emph{code-division multi-access} (CDMA) \cite{R33}. Building on these schemes, RRM is a broad area covering topics including admission control, scheduling, link adaptation, and interference coordination \cite{R34}. In this project, we consider two key topics in RRM, retransmission and scheduling, for their importance in practical systems e.g., \emph{long-term evolution} (LTE).

{\bf Retransmission} is a simple method for ensuring communication reliability in the presence of fading \cite{R6}. Its principle is to repeat the transmission of a data packet until it can be correctly decoded at a receiver. Thus, a retransmission technique essentially performs dynamic time allocation to transmission of different packets under the reliability constraint. Among different variations, a particular retransmission technique of our interest is called Hybrid \emph{automatic repeat request} (ARQ) implemented in LTE systems. Its key feature is to combine different transmitted versions of the same packet by \emph{maximal-ratio combining} (MRC) so as to incrementally enhance its receive \emph{signal-to-noise ratio} (SNR).

Focusing on radio access, traditional scheduling algorithms attempt to balance two goals. One is to allocate radio resources to meet different users' QoS requirements, namely their desired data rates and latency requirements \cite{R35}. The other goal is to maximize the spectral efficiency. Scheduling driven by this goal attempts to exploit \emph{multiuser diversity}, referring to independent fading in different users' channels \cite{R36}. However, this causes biased resource allocation in favor of users with best channels and being unfair to others. Practical algorithms usually attempt to balance two goals by maximizing spectral efficiency while observing some fairness constraints. Examples include proportionally fair scheduling \cite{R37} or max-min fair scheduling \cite{R38}. 

\begin{figure*}[h]
\begin{center}
{\includegraphics[width=15cm]{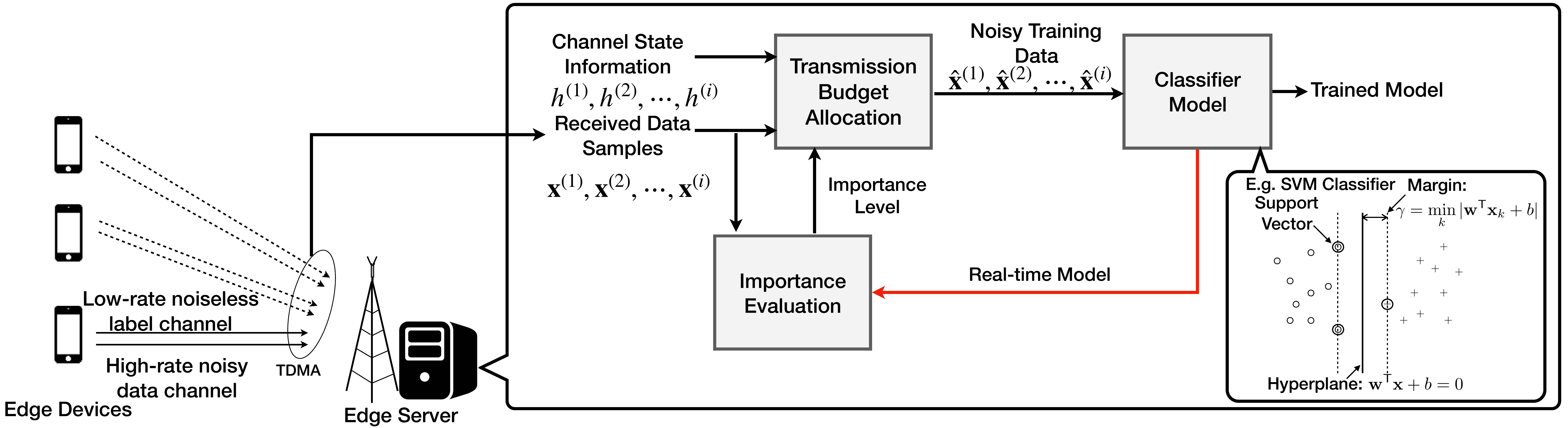}}\\
\vspace{-5pt} 
\caption{An edge learning system.}
\vspace{-25pt} 
\label{fig:3}
\end{center}
\end{figure*}

One can see that traditional RRM designs targeting radio-access networks are mainly driven by spectral efficiency and reliability. On the other hand, importance-aware RRM targeting edge-learning systems has learning performance as the main concern as illustrated in Fig. \ref{fig:3}. This motivates the consideration of data importance in the design, yielding new techniques such as importance-aware retransmission and scheduling as introduced in the sequel. 

While conventional RRM targets radio access, the new class of importance-aware RRM techniques share the following principle (also see Fig. \ref{fig:3}).
\begin{framed}
\noindent{\bf Principle 1} (Principle of importance-aware RRM for edge learning). In an edge-learning system, radio resources should be allocated to edge devices for transmissions based on the \emph{importance levels} of their local data and their \emph{channel states}.
\end{framed}

Thereby, data importance analytics provides a new dimension for improving the communication efficiency and coping with the negative effect of wireless propagation on learning performance. Such an approach has been adopted in practice such as Google's implementation of federated learning systems. In the following sections, we shall discuss specific research directions of importance-aware RRM for centralized edge learning and federated edge learning, as summarized in Table 1.

\begin{figure*}[h]
\begin{center}
{\includegraphics[width=17cm]{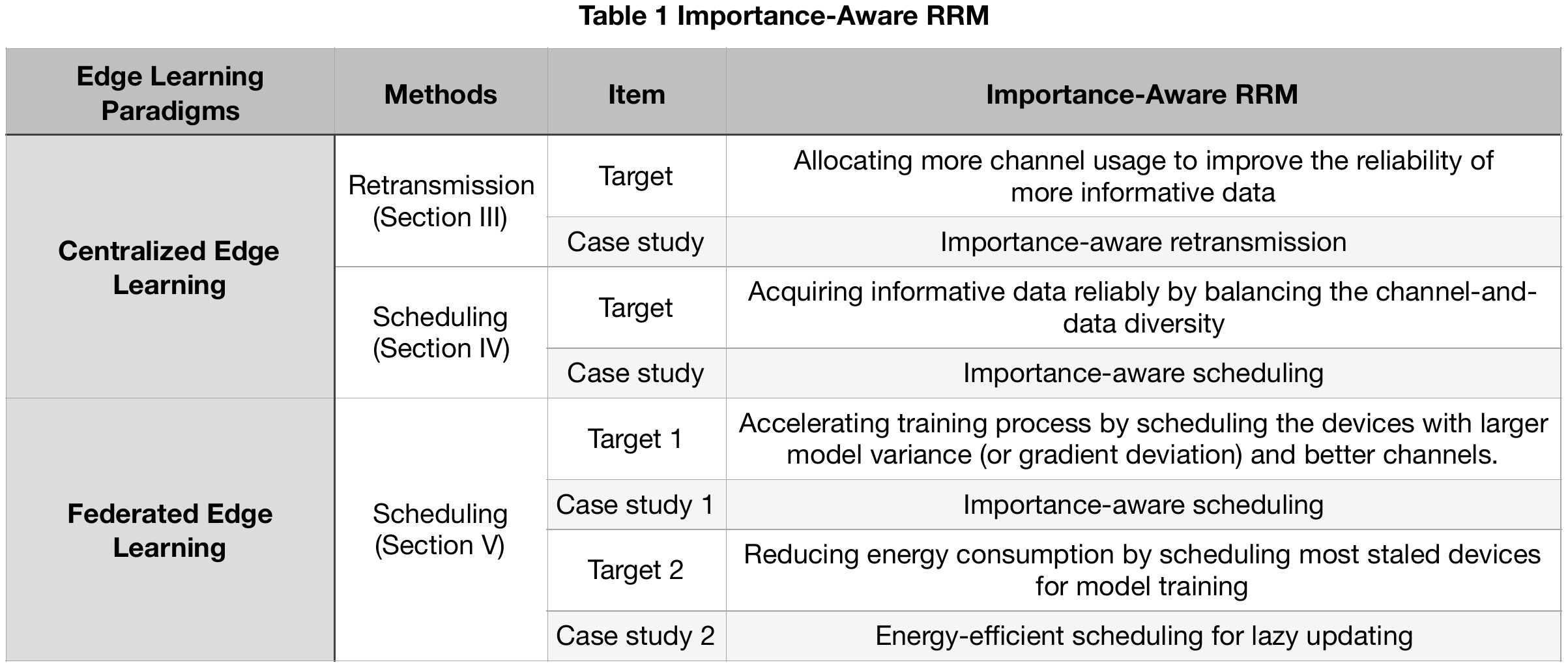}}\\
\end{center}
\end{figure*}


\section{Importance-Aware Retransmission}
\subsection{Principle and Opportunities}
Importance-aware retransmission are new retransmission protocols for active data acquisition in centralized edge learning systems. Designing the optimal retransmission policies need address two issues:
\begin{itemize}
\item The {\bf importance distribution} over data samples are non-uniform. The distribution affects the optimal time allocation for transmitting individual samples.

\item {\bf Quality-vs-quantity tradeoff.} Increasing the average number of retransmission can improve quality of the acquired dataset but reduce its size (quantity) due to a finite budget, giving rise to the said tradeoff. Though both quality and quantity of the training dataset are important for accurate learning, they need be balanced given a limited transmission budget (time slots).
\end{itemize}

By addressing these issues, there are several promising research opportunities that warrant future investigation.
\begin{itemize}
\item[1)] {\bf Uncertainty-aware ARQ:} The technique aims at optimizing the said quality-and-quantity tradeoff by controlling the retransmission of each data sample based on its uncertainty and effective SNR (after MRC). The key design step is the mathematical derivation of the tradeoff between the effective SNR and data uncertainty under a constraint on learning accuracy. The uncertainty can be measured using either entropy or distance to classification boundary. A key design challenge is to derive a tractable measure of learning accuracy. In the subsequent design example, the measure therein is derived based on ensuring a high likelihood that the received data sample agrees with its label, regulating the effect of channel noise on learning. Building on the derived SNR-uncertainty tradeoff, a retransmission policy can be then designed to enable intelligent data acquisition for accelerating AI-model convergence.

\item[2)] {\bf Batch-mode ARQ exploiting local data diversity:}  This direction targets the scenario of batch-mode training where data is acquired in batches (in e.g., a broadband or MIMO system) and the AI-model is updated using batch data (instead of single samples as in the preceding sub-task). Consider data pre-selection at an edge device where a batch is selected from the local dataset for transmission. Besides uncertainty, it is important to consider the diversity of the samples within the same batch as discussed earlier. Both metrics can be integrated in a single measure called Fisher information matrix \cite{R39}. Then {\bf uncertainty-and-diversity aware ARQ} can be designed by mathematically deriving a tradeoff between effective SNR and Fisher information.  Building on the result, it is possible to design practical retransmission policies for centralized edge learning.

\item[3)] {\bf Importance-aware ARQ exploiting global data diversity:} Importance-aware retransmission can be also designed to account for global data diversity. Specifically, data to be acquired (single sample or a batch) should be informative w.r.t. to the global dataset at the edge server. In particular, for classification, it is highly desirable to obtain new data guiding a classifier model to explore a new dimension with high discriminant gain but insufficient existing data. Based on this principle, policies on making retransmission decisions can be designed to account for global diversity of received data samples, which can be evaluated at the server using the global dataset and classifier model. This requires modification of the mentioned Fisher information matrix and then can leverage a similar design approach as described in the last bullet. 

\item[4)] {\bf Extension to federated edge learning:} The preceding techniques designed for active data acquisition can be extended to the paradigm of federated edge learning. The designing principles are similar while specific techniques need be redesigned due to the changes on the learning approach and importance metrics (to model variance and gradient divergence). 
\end{itemize}

\subsection{Example Design of Importance-Aware ARQ}
Based on the recent work in \cite{R40}, a design example of importance-aware ARQ is introduced as follows. 

\noindent{\bf System model:} Consider the edge learning system with retransmission as shown in Fig. \ref{fig:3}. There are multiple edge devices and an edge server, each equipped with a single antenna. The labelled dataset distributed over devices is transmitted to the server to train a classifier model. The server and the devices are connected by two separate channels: a low-rate noiseless label channel and a high-rate data channel with noise and fading. Upon receiving a data sample, the edge server makes a binary decision on whether to request a retransmission to improve the sample quality or a new sample from the scheduled device. The decision is communicated to the device by transmitting either a positive ACK or a negative ACK. Upon receiving a request from the server, each device transmits either the previous sample or a new sample randomly picked from its buffer. During each symbol block, the active device sends the data using linear analog modulation that is found recently to be an efficient method for multimedia transmission \cite{R41}. To reduce the overhead of broadcasting a large-size model to edge devices, the model can be compressed by pruning the model parameters with small values \cite{R40}. Unlike inference, the use of model for importance evaluation can tolerate aggressive compression without incurring significant performance degradation.

To exploit the time-diversity gain provided by retransmitting, multiple noisy observations of the same data sample are combined coherently using the MRC to maximize the receive SNR. The objective of designing the retransmission protocol is to minimize the duration of wireless data acquisition, thereby reducing radio-resource consumption and accelerating learning.

\noindent{\bf Learning model:} The supervised training of a classifier model at the server is considered. A coarse model at the server is refined progressively in the process of data acquisition plus training. A classical \emph{support vector machine} (SVM) classifier is considered in this example design as shown in Fig. \ref{fig:3}, which seeks an optimal hyperplane as a decision boundary by maximizing its margin to data points, i.e., the minimum distance from the hyperplane to each data point \cite{R42}. The points defining the margin, namely support vectors, directly determine the decision boundary (see Fig. \ref{fig:3}). The importance of a data sample for learning is measured by its uncertainty to the classifier. Since a SVM classifier makes less confident inference on a data sample which is located near the decision boundary, the uncertainty of a data sample can be suitably measured by the inverse of its distance to the boundary.

\begin{figure}[h]
\begin{center}
{\includegraphics[width=8.5cm]{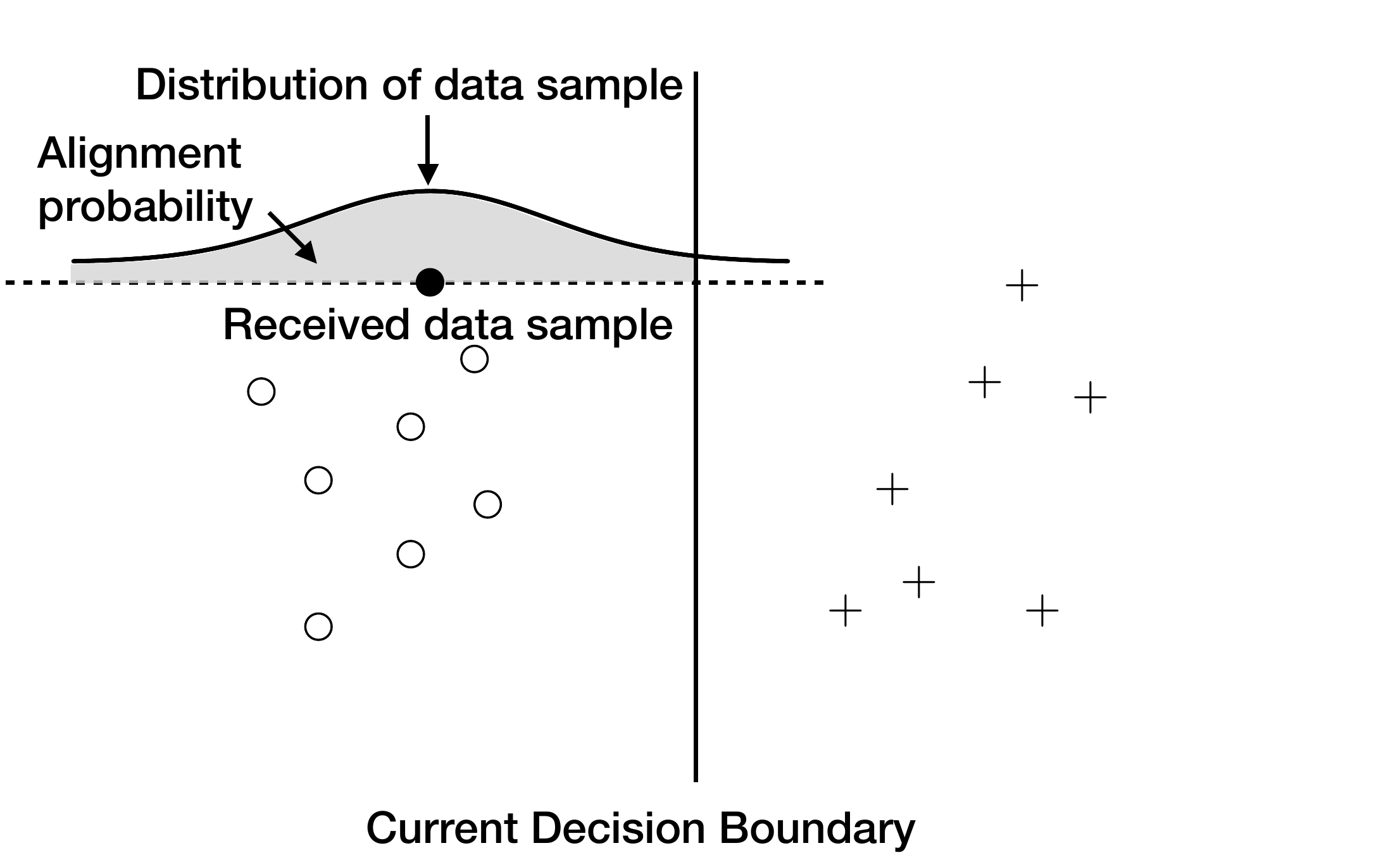}}\\
\caption{Illustration of noisy data alignment and data-alignment probability.}
\vspace{-20pt} 
\label{fig:4}
\end{center}
\end{figure}

Since a noisy received data sample may mislead the model training if the predicted label differs from the ground truth received without noise, a pair of transmitted and received data samples should be forced to lie at the same side (ground-truth) of the decision boundary in the classifier model so that the predicted labels are identical. This event is called noisy data alignment and its probability is referred to as the {\bf data-alignment probability} \emph{as illustrated in Fig. \ref{fig:4}}. From the distance-based uncertainty definition for SVM, the data samples with higher uncertainty are closer to the boundary and thus more vulnerable to noise corruption.

\noindent{\bf Importance-aware ARQ design for binary classification:} The ARQ protocol is designed to bias channel-use allocation towards providing better protection for more important (nearer to the SVM classification boundary) data samples against channel noise. To provide a guarantee on model accuracy, the design satisfies a constraint on the data-alignment probability defined early. This leads to a varying receive-SNR constraint on a data sample based on its importance level. The importance-aware ARQ protocol is a threshold based control policy with a SNR threshold adapted to data importance given as below.
\begin{framed}
\noindent{\bf Protocol 1} (Importance-aware ARQ for binary SVM classification). Consider the acquisition of a data sample ${\bf x}$ from a scheduled edge device. The edge server repeatedly requests the device to retransmit ${\bf x}$  until the effective receive SNR satisfies
\begin{equation}\label{eq:1}
{\rm SNR} > \min ( \theta_{0} \; \mathcal{U}_{\sf d}(\bf x), \theta_{\rm SNR}),
\end{equation}
where $\theta_0$ represents the data-alignment probability, $ \mathcal{U}_{\sf d}(\bf x)$ represents the distance-based uncertainty, and $\theta_{\rm SNR}$ is a given maximum SNR.
\end{framed}
It can be observed from \eqref{eq:1} that the SNR threshold  is  proportional  to  the  distance-based uncertainty of the data sample. One can see in Fig. \ref{fig:4} that a sample near the decision boundary has a higher level of uncertainty (importance) but is more vulnerable to noise corruption. Thus, it requires a higher receive SNR and hence more retransmissions to achieve the required data-alignment probability. This is aligned with the result in \eqref{eq:1}.

Next, the learning performance of the proposed importance-aware ARQ is evaluated by experiments on handwritten-digit recognition using the well-known MNIST dataset. The dataset consists of 10 categories ranging from digit ``0" to ``9" with a total of 60,000 labeled training data samples. In the experiments, two relatively less differentiable classes of ``3" and ``5" letters are chosen for binary classification using SVM. The maximum transmission budget is set to be 4,000 (channel uses). Two baseline protocols are considered. One is without retransmission (maximizing the size of acquired data) and another is the conventional channel-aware ARQ (in e.g., LTE), where the retransmission decisions depend merely on the channel states.

\begin{figure*}[h]
\begin{center}
{\includegraphics[width=13cm]{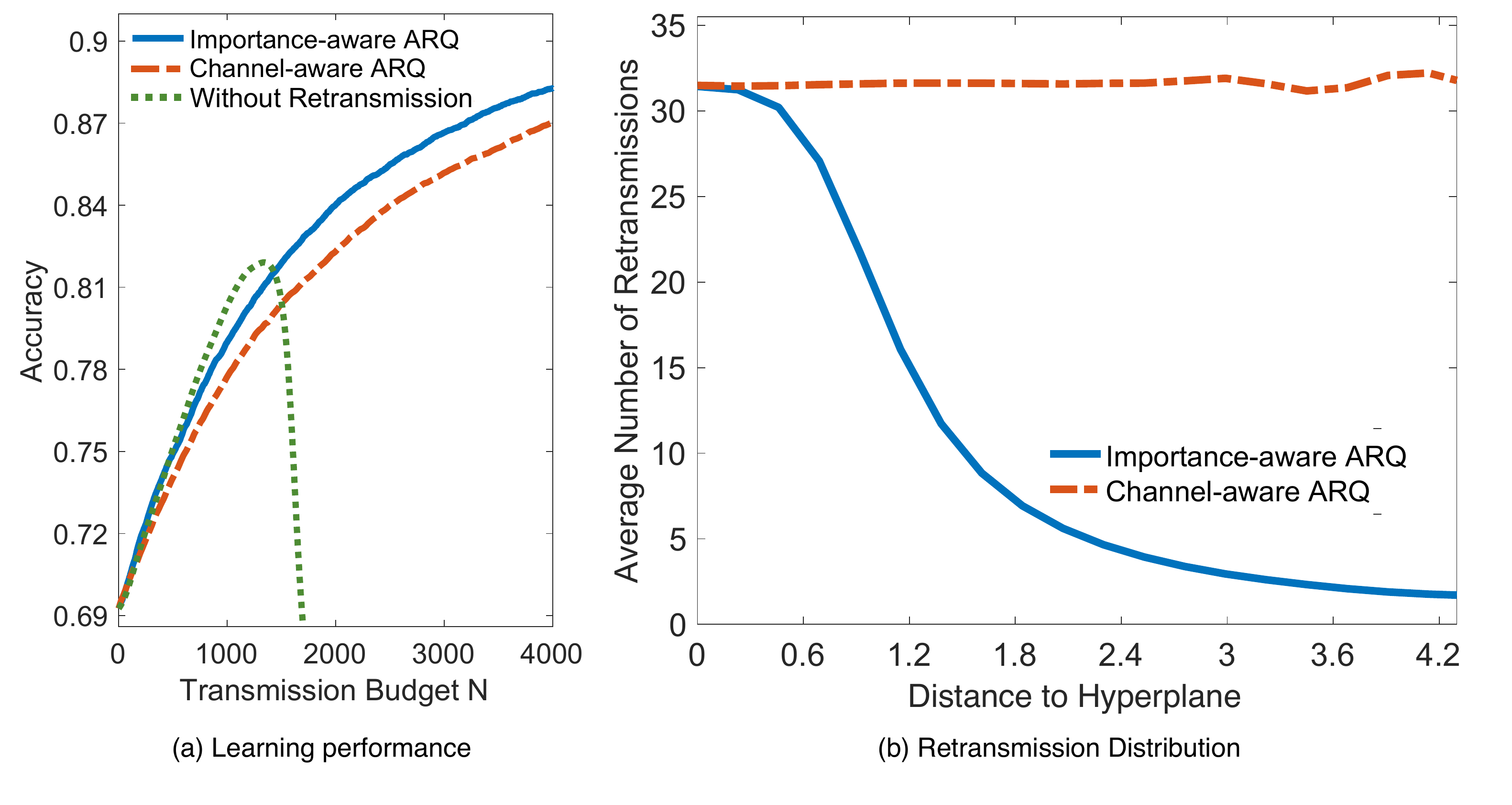}}\\
\vspace{-10pt} 
\caption{Learning performance for a binary SVM classifier trained with wirelessly acquired data.}
\vspace{-25pt} 
\label{fig:5}
\end{center}
\end{figure*}

 As illustrated in Fig. \ref{fig:5}(a), the performance of the protocol without retransmission degrades dramatically when the number of received samples is sufficiently large as they are too noisy. This emphasizes the need of retransmission. On the other hand, the importance-aware ARQ outperforms the channel-aware ARQ throughout the whole training duration. This confirms the performance gain due to taking into account of data importance in radio resource allocation for data acquisition. The underlying reason for the performance improvement of importance-aware ARQ is further illustrated by plotting the distribution of average numbers of retransmissions over a range of sample uncertainty (inversely proportional to sample distance to the decision boundary) in Fig. \ref{fig:5}(b). It can be observed that the channel-aware ARQ has close-to-uniform distribution since the SNR is independent on the sample uncertainty. In contrast, the retransmission for importance-aware ARQ is concentrated in the high uncertainty region. This is aligned with the importance-aware design principle.
 
The extension of the design to convolutional neural network models can be found in \cite{R40} where significant performance gain of importance-aware design is also observed. 

\noindent{\bf Importance-aware ARQ  design for multi-class classification:} In this section, the importance-aware ARQ developed for binary classification is extended for multi-class classification.  By using the one-versus-one implementation \cite{R43}, a $C$-class SVM classifier can be decomposed into $L = C(C-1)/2$ binary component classifiers, each is trained by the data samples of the two concerned classes. The corresponding output of each input sample ${\bf x}$ is a $L$-dimension vector recording the $L$ scores, denoted by $\bold{s} = [s_1(\bold{x}),s_2(\bold{x}), \cdots, s_L(\bold{x})]$. A reference coding matrix $\bold{M} \in \mathbb{R}^{C \times L}$ is built to map the output ${\bf s}$  to the class indexes. By comparing the Hamming distances between ${\bf s}$ and each row in ${\bf M}$, the predicted class index $\hat{\bf c}$ is the row index with the smallest distance. Consequently, the single-threshold policy for importance-aware ARQ is extended to a multi-threshold policy given as below.
\begin{framed}
\noindent{\bf Protocol 2} (Importance-aware ARQ for multi-class SVM classification). For training a $C$-class SVM classifier, the edge server repeatedly requests the device to retransmit ${\bf x}$ until the effective receive SNR satisfies
\begin{equation}
\rm{SNR} > \min (\theta_{0}/|s_\ell(\bold{x})|^2, \theta_{\rm{SNR}}),~ \forall \ell \in \{\ell| m_{\hat{c}\ell} \neq 0\},
\end{equation}
where $m_{\hat{c}\ell}$ represents the element in the $\hat{c}$-th row and the $\ell$-th column of ${\bf M}$, $|s_\ell(\bold{x})|^2$ represents the distance-based uncertainty of the $\ell$-th binary component classifier.
\end{framed}

As in the binary SVM case, experimental results show that the importance-aware ARQ  consistently outperforms the two conventional schemes, namely the channel-aware ARQ scheme and the non-retransmission scheme.


\section{Importance-Aware Scheduling for Centralized Edge Learning}
\subsection{Principle and Opportunities}
While retransmission focuses on RRM in time, scheduling is another dimension to explore for importance-aware RRM in a multiuser edge-learning system. There exist two types of multiuser diversity in such a system. One is the independent fading in multiuser channels, called \emph{multi-user channel diversity}. The other is the heterogeneous importance levels across multiuser datasets, called \emph{multi-user data diversity}. Conventional scheduling in wireless networks exploits only the former. However, in edge learning system, it is desirable to jointly exploit both types of diversity so as to maximize the communication efficiency. However, doing so leads to two potentially conflicting goals: one is to schedule devices with the best channels to maximize the data-transmission reliability and the other is to select devices with most informative data. Since both goals are important for learning, they need to be balanced. This sets the following objective of designing \emph{importance-aware scheduling}.
\begin{framed}
\noindent{\bf Principle 2} (Principle of importance-aware  scheduling for centralized  edge learning). Targeting active wireless data acquisition for centralized learning, the importance-aware scheduling should simultaneously exploit the {\bf multi-user channel-and-data diversity} to improve edge-learning performance.
\end{framed}

Based on the above principle, we can envision the rise of many interesting research opportunities with some described as follows.
\begin{itemize}
\item[1)] {\bf Joint channel-and-uncertainty scheduling:} A key step of designing the scheduling algorithm is to derive a \emph{data-importance indicator} (DII) as a function of both data uncertainty and reliability (measured by e.g., channel state) as shown in Fig. \ref{fig:6}. For instance, an indicator can be the uncertainty of a data sample (evaluated using the real-time model under training) under a constraint on the data-alignment probability defined in the preceding section for providing a guarantee on learning performance. Then maximizing the DII can be applied as a criterion for scheduling to ensure both the data's reliability and usefulness for learning. 

\begin{figure}[h]
\begin{center}
{\includegraphics[width=8.5cm]{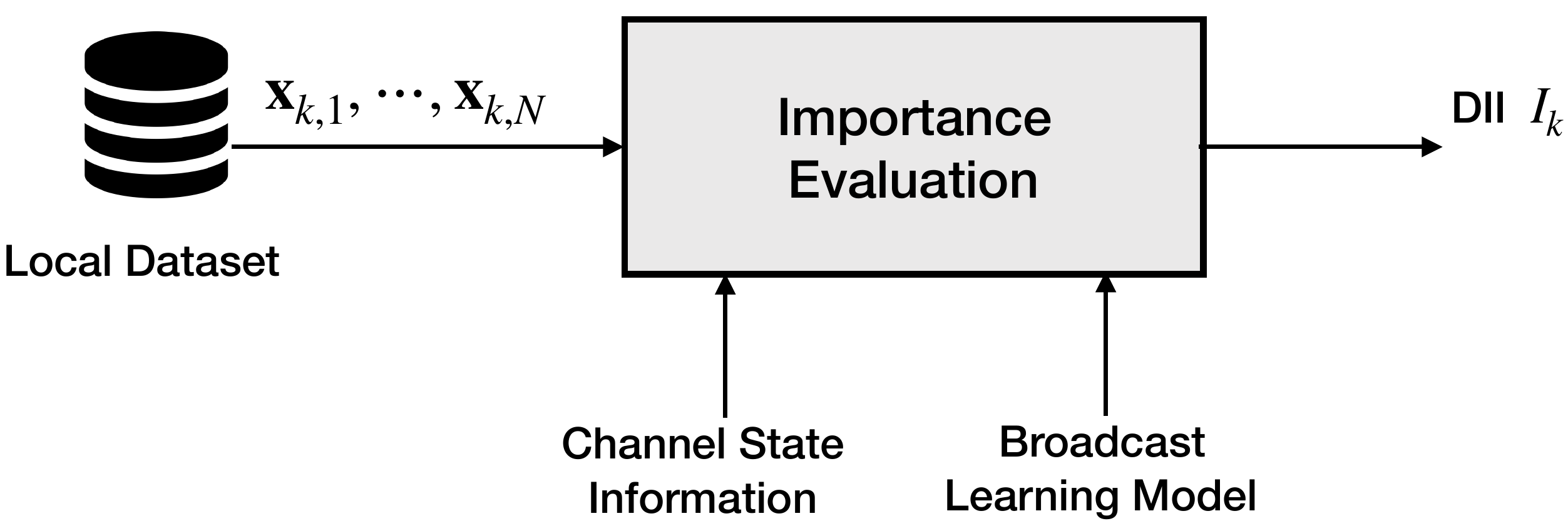}}\\
\caption{The computation of data-importance indicator at an edge device.}
\vspace{-10pt} 
\label{fig:6}
\end{center}
\end{figure}

\item[2)] {\bf Batch-mode scheduling exploiting multiuser data diversity:} Consider a broadband system with OFDMA or a MIMO system with SDMA, multiple devices can be scheduled for simultaneous transmission and thereby distributed data samples can be acquired in batches. This allows more efficient training of the AI model in the batch mode. For retransmission in the preceding section, it is desirable to ensure data diversity in each batch besides uncertainty of individual samples. The current scenario is more complex as it concerns multiuser data diversity instead of local diversity at a single device in the retransmission scenario. Correspondingly, a sophisticated DII function for the current scenario need be derived to account for data diversity, uncertainty, and reliability. The function can build on the mentioned Fisher information matrix as an integrated measure of data uncertainty and diversity. Subsequently, by developing a suitable signalling procedure, the derived DII can be applied to scheduling for edge learning in the batch mode. 

\item[3)] {\bf Scheduling exploiting global data diversity:} The scheduling algorithms can be further designed to ensure global data diversity introduced earlier. This requires redesigning the DII function to account for both multiuser diversity and global diversity.
\end{itemize}

In the sequel, an example design is discussed for importance-aware scheduling for centralized edge learning systems.

\subsection{Example Design of Importance-Aware Scheduling}
\begin{figure}[h]
\begin{center}
{\includegraphics[width=9cm]{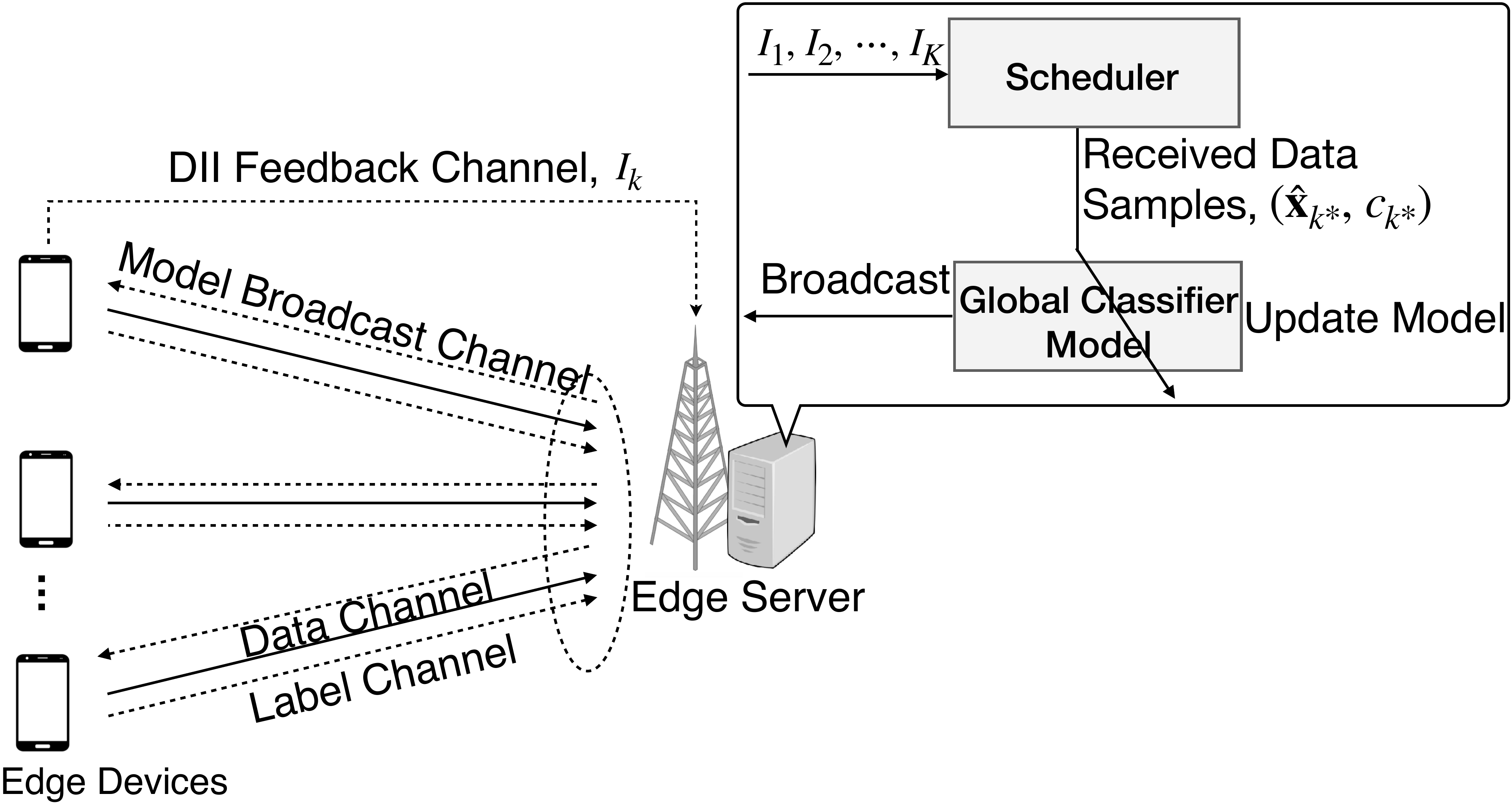}}\\
\caption{An edge learning system with importance-aware scheduling.}
\vspace{-20pt} 
\label{fig:7}
\end{center}
\end{figure}

\noindent{\bf System model:} Consider a multiuser edge learning system with one server and $K$ devices, as shown in Fig. \ref{fig:7}. Each device is equipped with a local buffer storing $N$ data samples.  To train a classifier model at the server, the devices time share the wireless channel to upload the training data samples to the server. In each sample block, there are three steps for selecting a device to upload one data sample. First, the edge server broadcasts the current global model to all devices. Second, the data importance is evaluated at each device, measured by the DII and denoted as $I_k$ for the $k$-th device. Finally, the device with largest DII is selected for data uploading. As in the preceding section, analog modulation is adopted for data transmission. The received data sample at the server is distorted by channel fading and noise. The distortion depends on the SNR, which can be calculated at both the server and devices, all of which have perfect CSI.

\noindent{\bf Learning model:} As in the preceding section, binary soft-margin SVM model is considered. Prior to wireless data acquisition, the server has a coarse initial classifier that can be used for data-importance evaluation at the beginning and will be refined progressively in the training process. The importance of a noisy received data sample can be measured using its \emph{expected uncertainty}. Based the distance-based uncertainty measure introduced in the preceding section, the expected uncertainty is translated to the expected distance between the data sample and the current decision boundary of the classifier model.

\begin{figure}[h]
\begin{center}
{\includegraphics[width=8.5cm]{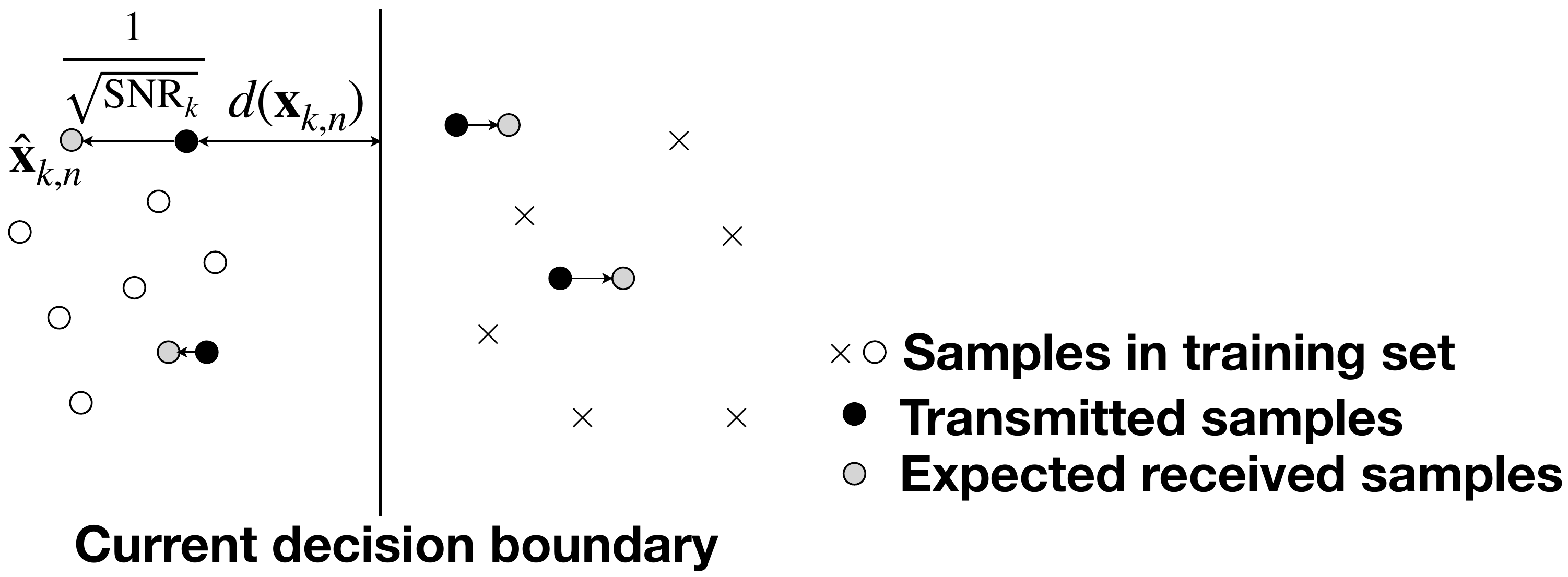}}\\
\caption{Effect of channel fading and noise on DII with unlabelled-data.}
\vspace{-15pt} 
\label{fig:8}
\end{center}
\end{figure}

\noindent{\bf Importance-aware scheduling:} Targeting the scenario where labelling is costly, the data samples in the devices are assumed to be unlabelled. And the label is generated for the selected data after transmission to the server by recruiting a labeller. To design the scheduling scheme, the effect of channel fading and noise on the expected received data samples at the server is first analyzed, as illustrated in Fig. \ref{fig:8}, where ${\rm SNR}_k$ and ${\bf x}_{k,n}$ are the SNR and the $n$-th data sample of the $k$-th device, and $d({\bf x}_{k,n})$ is the distance between ${\bf x}_{k,n}$ and the decision boundary, respectively. The DII of the selected data sample in the $k$-th device, refers to the sample with the maximum expected uncertainty among the local data. In \cite{R44}, the DII is shown to have the following expression
\begin{equation}\label{eq:3}
I_k= -\frac{1}{{\rm SNR}_k}+\max_{n\in\mathcal{N}_k} {\mathcal{U}_{\sf d}\left({\bf x}_{k,n}\right)},
\end{equation}
where $\mathcal{N}_k=\{1,2,\cdots,N\}$ represents the sample index set, and $\mathcal{U}_{\sf d}\left(\cdot\right)$ is a distance-based uncertainty measure. One can observe that in the derived DII, both data uncertainty (the last term) and the channel quality (the SNR term) are in a simple addition form. The DII being a monotone increasing function of SNR is due to the fact that the channel fading and noise tend to degrade the importance of data samples by making their distances to the decision boundary more likely to be larger than smaller (see Fig. \ref{fig:8}). In other words, channel distortion harms the learning performance which is aligned with our intuition.

Based on the DII in \eqref{eq:3}, the scheduling scheme for binary SVM is to select the device with largest DII in each sample block, as described in Scheme 1.
\begin{framed}
\noindent{\bf Scheme 1} (Importance-aware scheduling without label information). Consider the acquisition of a data sample from multiple edge devices in an edge learning system. The edge server schedules device $k^*$ for data transmission if 
\begin{equation}
k^*=\arg \max_{k}\;\left\{-\frac{1}{{\rm SNR}_k}+\max_{n\in\mathcal{N}_k} {\mathcal{U}_{\sf d}\left({\bf x}_{k,n}\right)} \right\}.
\end{equation}
\end{framed}
The design can be easily extended to a general classifier by replacing the distance-based uncertainty measure with a general measure. The criterion in Scheme 1 shows that simultaneous exploitation of multi-user data-and-channel diversity is required for learning performance improvement rather than a single type of diversity. Besides, the influence of the wireless transmission, i.e., the transmit SNR, on the scheduling criterion is interesting. When the wireless channels are unreliable, saying the SNR is low, the received data samples at the server are severely corrupted by channel noise and become useless regardless of their uncertainty (importance) before transmission. In this case, only multi-user channel diversity is exploited. When the SNR is large, it is more critical to exploit the data diversity, as there is little distortion of the received data samples.

The extension of Scheme 1 to \emph{convolutional neural network} (CNN) models can be found in \cite{R44} by essentially replacing the uncertain measure with one that fits CNN, e.g., entropy.

\noindent{\bf Performance:} Experiments are carried out to verify the performance gain of the importance-aware scheduling with respect to conventional schemes exploiting only a single type of multiuser diversity. The first benchmarking scheme, namely channel-aware scheduling, only utilizes the multiuser channel diversity and the other, namely data-aware scheduling, only exploits the multiuser data diversity. The experimental settings are as follows. There are $K=10$ edge devices in the system, each of which is equipped with a local buffer with the size $N=10$. The transmission budget $T$ for the binary SVM  learning task is 100 channel uses, each of which is for transmitting a single data sample. Rayleigh fading channels with unit variance are considered with the average transmit SNR=15 dB. The well-known MNIST dataset also used in previous experiments is adopted for training.

\begin{figure}[h]
\begin{center}
{\includegraphics[width=7.5cm]{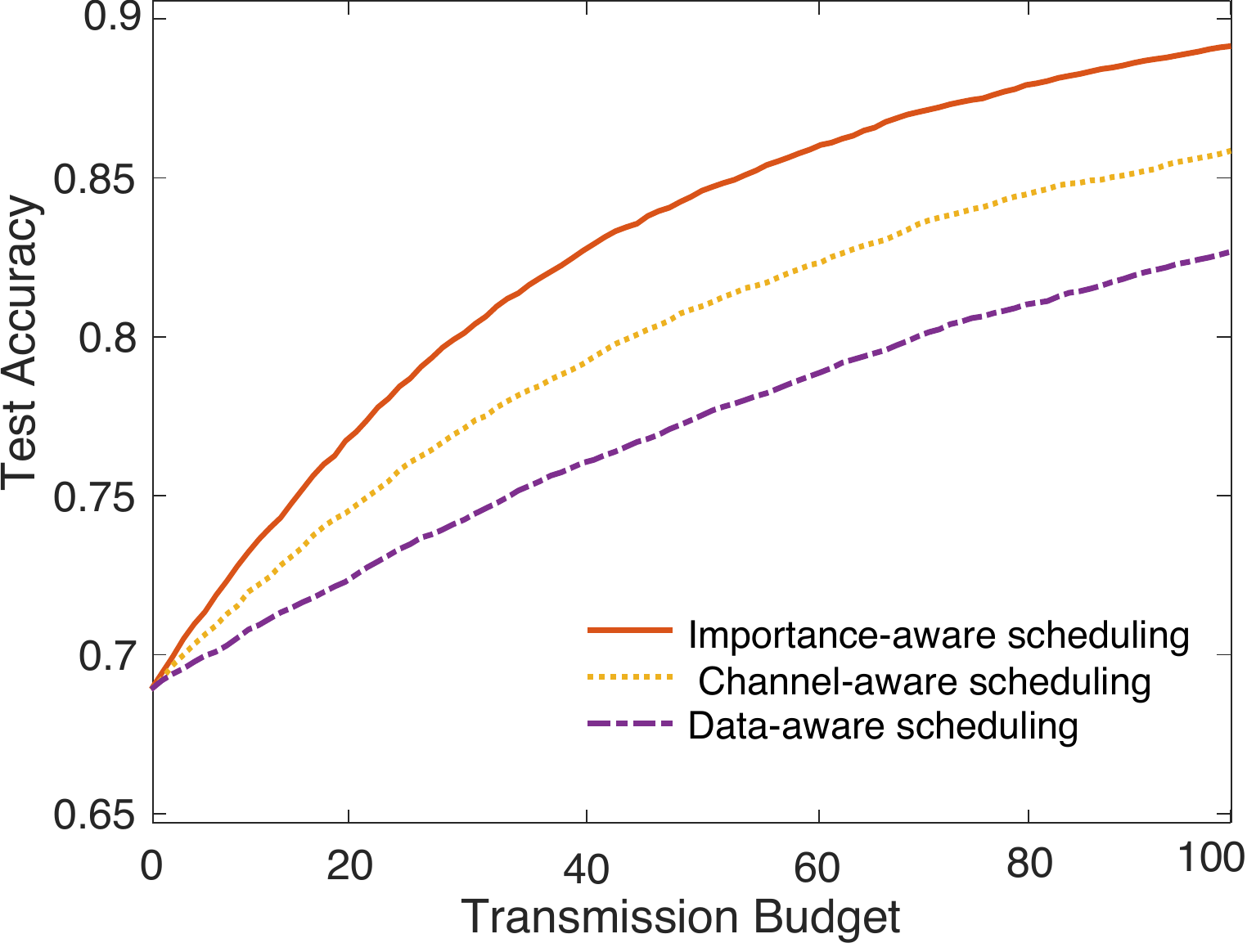}}\\
\caption{Learning performance evaluation for importance-aware scheduling.}
\vspace{-15pt} 
\label{fig:9}
\end{center}
\end{figure}

The test accuracies of models trained using different schemes are compared in Fig. \ref{fig:9}. One can observe that importance-aware scheduling can achieve signifiant improvement in test accuracy of about 5\% over channel-aware scheduling and of about 8\% over data-aware scheduling. Moreover, the model convergence of the new design is faster too.

\section{Importance-Aware Scheduling for Federated Edge Learning}
\subsection{Principle and Opportunities}
The previous designs of importance-aware scheduling for centralized edge learning can be extended to the paradigm of federated edge learning. The extension essentially involves changing the data-importance metrics to model variance and gradient divergence that suits federated learning. On the other hand, scheduling designs should also factor in some unique features of federated learning such as the following two.
\begin{itemize}
\item {\bf Update synchronization:} The key operation of averaging local updates (local gradients/models) at the server requires synchronization of updates by edge devices. To be specific, the duration of one communication round is determined by the slowest device in computing and communication. As a result, the learning latency can be excessive when the number of devices are many and they have high heterogeneity in computing capacities, channels, and local dataset sizes.
\item {\bf Communication-computation tradeoff and lazy updating:} Researchers have discovered a \emph{communication-computation tradeoff} that increasing local computation load can be traded for reduced communication overhead \cite{R45}. To be specific, increasing the number of rounds for only updating local models at devices allows the reduction of communication frequency for updating the global models, called ``lazy updating", without incurring significant performance loss.
\end{itemize}

Based on the above discussion, a design principle of importance-aware scheduling for federated edge learning can be proposed as follows.
\begin{framed}
\noindent{\bf Principle 3} (Principle of importance-aware scheduling for federated edge learning).
Targeting active wireless data acquisition for federated learning, the importance-aware scheduling should exploit 1) the {\bf multi-user diversity in channels and local updates (channel-and-data diversity)} and 2) {\bf communication-computation tradeoff} under the constraint of update synchronization to improve edge-learning performance. 
\end{framed}

The fundamental changes on the scheduling principle compared with the conventional rate-maximization paradigm introduce many new challenges and give rises to many research opportunities. A few are described as follows.

\begin{itemize}
\item[1)] {\bf RRM for update synchronization:}  Due to the mentioned heterogeneity among edge devices, there exist \emph{``stragglers"} (slow devices) among them that can potentially slow down the learning process. To alleviate the bottleneck, one way is to design scheduling for allocating more bandwidth for transmission by stragglers and less for fast devices \cite{R46}. This to some extent can equalize their total latency (computing plus communication latency) and thereby facilitate update synchronization needed for federated learning. On the other hand, to avoid squandering bandwidth on extremely slow devices, scheduling should exclude slowest devices by applying thresholds on their computing capacities or channel capacities or both. Note that more updating devices lead to faster model convergence (in terms of the number of communication rounds) by exploiting a larger distributed dataset. Hence the said thresholds should balance the two conflicting effects of increasing the number of scheduled devices: 1) reducing the number of communication rounds and 2) increasing the computation-and-communication latency per round. In summary, RRM (e.g., scheduling and bandwidth allocation) can be designed by optimizing the number of scheduled devices based on the tradeoff to reduce the overall learning duration in seconds.

\item[2)] {\bf Solving the dilemma in lazy updating:} There exists a dilemma for lazy updating. Being too lazy in updating of the global AI model will reduce the average number of updating devices in each round and thereby compromise the learning performance. That, however, will reduce the total communication overhead (and hence latency) in each round. This creates a dilemma that will be solved by scheduling design. This requires a quantitative understanding of the dependence of the learning accuracy and latency (in seconds but not the number of communication rounds) on a few factors including the level of updating laziness, channel distortion and latency, number of devices, and the amount of radio resources. The task will require the interplay of communication theory and federated learning theory (e.g., SGD convergence analysis). The results will be applied to improve the previously developed algorithms of importance-aware scheduling to integrate the optimal level of laziness to minimize communication overhead. Thereby, communication latency for federated edge learning can be further reduced.
\end{itemize}

\subsection{Sketch of Two Example Designs}
There exist few results on importance-aware scheduling for federated edge learning that integrate RRM and learning. To help readers to better understand earlier discussion, we sketch two example designs with some concrete steps delegated to future investigation.

\subsubsection{Importance-aware scheduling for exploiting multiuser update diversity}  As mentioned, for federated learning, a local gradient with a larger norm or a local model with a larger variation (w.r.t. the global model) tend to contribute more significantly to the global model convergence \cite{R24}. In particular, the global loss function is observed to decrease approximately with the square of the global gradient norm, on which local gradients with larger norms have more influences. This justifies the use of gradient norm or model variation as an importance metric for scheduling. The aim of the scheduling design is to simultaneously exploit multiuser diversity in both channels and the local updates (gradient/model). The scheduling procedure and the open challenge are described as follows. 

\begin{itemize}
\item {\bf Step 1 (Global Model Broadcasting):} At the beginning of each communication round, the edge server broadcasts the current global model to all devices.

\item {\bf Step 2 (Local Gradient/Model Calculation):} Based on the received model, each device uses local data to compute an updated local model (or a gradient for updating the model) using the backpropagation algorithm. Subsequently, the gradient norm (or model variation) can be computed.

\item {\bf Step 3 (Importance Indicator Uploading):} Each device reports the scalar importance indicator (gradient norm or model variation) and CSI to the server via some control channel.

\item {\bf Step 4 (Importance-Aware Scheduling):} By collecting the importance indicators and CSI from all devices, the edge server selects a subset of devices for uploading and allocates radio resources (bandwidth or time) for their transmission. Designing the scheduling policy can be formulated as a joint optimizaiton problem over user selection and bandwidth allocation for the goal of improving learning performance. Some suitable criteria can be maximizing the model test accuracy under a latency constraint or minimizing the training latency under an accuracy guarantee. A tractable solution approach hinges on the derivation of learning performance as a closed-form function of local updates and channel states. This remains largely an {\bf open problem}.

\item {\bf Step 5 (Local Update Transmission):} The scheduled devices then transmit their concrete gradient vectors to the edge server for gradient aggregation and global model updating.
\end{itemize}
The above five steps will be iterated till the global model converges.

\subsubsection{Energy-efficient scheduling for lazy updating:} Computing a local gradient/model update is an energy consuming process as it involves repeated backpropagation over a CNN model typical having millions to billions of parameters. One issue with existing scheduling schemes for federated edge learning is that it requires all devices to perform such computation but only selects a subset for transmission (see e.g., \cite{R46}) ). This is not energy efficient when the number of devices is large. One idea to improve the efficiency is to limit the computation of gradient/model updates to other scheduled devices. Then implementing the idea requires the design of a new scheduling metric. Consider lazy updating discussed earlier. One suitable design is the variation between the current global model and the local model a device transmits a number of rounds ago and stores in its memory, called \emph{model age indicator} (MAI). To be specific, in the $i$-th round, assume that the local model stored in the device $k$ is $\hat{w}_k^{(i)}$ with staleness of $\tau_k^{(i-1)}$  rounds, i.e., $\hat{w}_k^{(i-1)}=w_k^{(i-1-\tau_k^{(i-1)})}$. Given the current global model $w^{(i)}$, the MAI  is given as $\lVert \hat{w}_k-w^{(i)} \rVert$. The key advantage of using the MAI as a scheduling metric is it is unnecessary for a device to perform model uploading, which is complex and energy hungry, unless it is scheduled for uploading. Instead, a device need only perform the less complex computation of the MAI using the stored model and received global model, and then report the MAI to the server. Thereby, the total energy consumption of devices are reduced.

In the context of lazy updating, the motivation of scheduling devices with relative MAI is that their data are infrequently explored in the global model training and thus scheduling them can yield significant model improvements due to data diversity. On the other hand, it is undesirable to keep MAI too small as too aggressive exploitation of multiuser data diversity loses the ``laziness" needed for exploiting multiuser channel diversity.

Based on the above discussion, the procedure for energy efficient lazy updating is as follows.
\begin{itemize}
\item {\bf Step 1 (Global Model Broadcasting):} The server broadcasts the global model to all devices.

\item {\bf Step 2 (MAI Reporting):} Each device computes the MAI using a stored local model and the received global model. Then all devices transmit their MAI to the server. 

\item {\bf Step 3 (Scheduling):} Using global MAI and CSI, the server schedules a subset of devices to upload the global model and allocates bandwidth for their transmission. The resource allocation policy can be formulated as an optimization problem for energy efficient  uploading based on their channel conditions and the time for transmission in a round. A tractable solution combines the energy consumption and learning performance as a closed-form function of channel conditions and computation capacities. This remains largely an {\bf open problem}.

\item {\bf  Step 4 (Local Model Updating and Transmission):} The scheduled devices update their local models and transmit them to the server using the allocated spectrums.

\item {\bf Step 5 (Global Update):} The edge server aggregates the gradients and updates the global model.
\end{itemize}
Again, the above steps will be iterated till the global model converges.

\section{Concluding Remarks}
Two 5G missions, namely gigabit access and tactile response (network response time of several milli-seconds), have not yet been fully realized.  On the one hand, the existence of massive number of subscribers and IoT devices congest the network and reduce the average access speeds. On the other hand, the latency of computing (at both devices and base stations), protocols (e.g., admission and routing), and round-trip wireless communication add up to multiply the total response time. The two missions will continue to drive the 6G development. As driven by the availability of massive mobile data, 6G will have a new mission of realizing ubiquitous computing and intelligence to support next-generation AI driven intelligent applications. This mission has resulted in the recent emergence of the new research area, edge AI. However, the full potential of edge computing and learning cannot be unleashed without gigabit access and tactile response. To tackle this challenge, the main approach of edge-AI research is to seamlessly integrate communication and learning theories. Importance-aware RRM represents a main thrust in this area. Advancements in this direction will make significant contributions towards realizing a communication-efficient intelligent network edge. Besides those discussed in the preceding sections, the importance-aware design principle can be also applied to other RRM techniques including spectrum allocation, power control, and multi-antenna transmission such that more bandwidth, power, and spatial degrees-of-freedom can be allocated to the uploading by devices with more important data.








\end{document}